\documentclass[final,3p,times,twocolumn]{elsarticle}
\usepackage{xcolor}

\journal{Physics Letters B}

\begin{document}

\begin{frontmatter}

\title{Proton-neutron pairing correlations in the self-conjugate nucleus $^{42}$Sc}


\author[1]{\'{A}.~Koszor\'{u}s}\ead{A.Koszorus@liverpool.ac.uk}
\author[1]{L.J.~Vormawah}
\author[2]{R.~Beerwerth}
\author[3]{M.L.~Bissell}
\author[3]{P.~Campbell}
\author[1]{B.~Cheal}
\author[1]{C.S.~Devlin}
\author[4]{T.~Eronen}
\author[2]{S.~Fritzsche}
\author[4,5]{S.~Geldhof}
\author[5,6]{H.~Heylen}
\author[7,8]{J. D.~Holt}
\author[4]{A.~Jokinen}
\author[3]{S.~Kelly}
\author[4]{I.D.~Moore}
\author[7]{T.~Miyagi}
\author[4]{S.~Rinta-Antila}
\author[4]{A.~Voss}
\author[1]{C.~Wraith}

\address[1]{Department of Physics, University of Liverpool, Liverpool L69 7ZE, United Kingdom}
\address[2]{Helmholtz Institute Jena, Fr\"{o}belstieg 3, 07743 Jena, Germany}
\address[3]{School of Physics and Astronomy, University of Manchester, Manchester M13 9PL, United Kingdom}
\address[4]{Department of Physics, University of Jyv\"{a}skyl\"{a}, PB 35(YFL) FIN-40351 Jyv\"{a}skyl\"{a}, Finland}
\address[5]{KU Leuven, Instituut voor Kern-en Stralingsfysica, B-3001 Leuven, Belgium}
\address[6]{CERN, Experimental Physics Department, CH-1211 Geneva 23, Switzerland}
\address[7]{TRIUMF, 4004 Wesbrook Mall, Vancouver, BC V6T 2A3, Canada}
\address[8]{Department of Physics, McGill University, 3600 Rue University, Montr\'eal, QC H3A 2T8, Canada}

\begin{abstract}
Collinear laser spectroscopy of the $N=Z=21$ self-conjugate nucleus $^{42}$Sc has been performed at the JYFL IGISOL IV facility in order to determine the change in nuclear mean-square charge radius between the $I^{\pi}=0^{+}$ ground state and the $I^{\pi}=7^{+}$ isomer via the measurement of the $^{42\mathrm{g},42\mathrm{m}}$Sc isomer shift. New multi-configurational Dirac-Fock calculations for the atomic mass shift and field shift factors have enabled a recalibration of the charge radii of the $^{42-46}$Sc isotopes which were measured previously. While consistent with the treatment of proton-neutron, proton-proton and neutron-neutron pairing on an equal footing, the reduction in size for the isomer is observed to be of a significantly larger magnitude than that expected from both shell model and ab-initio calculations.
\end{abstract}


\end{frontmatter}


Odd-odd self-conjugate ($N=Z$) nuclei provide an ideal testing ground for proton-neutron pairing studies. More specifically, such nuclei are ideal for verifying if the same phenomena arise for an $I=0$ proton-neutron ($\pi\nu$) pair as for a proton-proton ($\pi\pi$) or neutron-neutron ($\nu\nu$) pair with $I=0$ \cite{Frauendorf2014}. 
The charge independence of the nucleon-nucleon interaction suggests symmetry in isospin between protons and neutrons, though recent findings elucidate the violation of the mirror symmetry \cite{witek_mirror} in the ground state of bound nuclei on the edge of the nuclear landscape.

In odd-odd $N=Z$ nuclei, protons and neutrons occupy the same orbitals. Therefore, the interaction between the odd proton and neutron is enhanced, leading to a greater likelihood of the formation of a $\pi\nu$ pair. For such a nucleus, it is expected that the charge radius will be greater for a state with $I=0$, $T=1$ than for such with $I\neq 0$, $T=0$. This arises due to a similar orbital blocking picture to that accounting for the multi-quasiparticle isomer radii \cite{Bissell2007}. The $I=0$, $T=1$ (isovector) $\pi\nu$ pair, can scatter into a wide range of excited orbitals and still satisfy $I=0$; as these orbitals are less bound, they will have a greater spatial extent. The $I\neq 0$, $T=0$ (isoscalar) $\pi\nu$ pair, on the other hand, is significantly restricted in the number of orbitals it can mix with. In such cases, orbitals with the same spin may well be energetically far apart (possibly even across a major shell closure), rendering such scattering unlikely \cite{Bissell2014}.

A study of pairing effects on the relative mean-square charge radii of multi-quasiparticle isomers~\cite{Bissell2007}, such as $^{97\mathrm{m2}}$Y, $^{176\mathrm{m}}$Yb and $^{178\mathrm{m1}}$Hf, was motivated by the measurement of $^{178\mathrm{m2}}$Hf by Boos~\textit{et al.}~\cite{Boos1994}. All of these isomers were found to have a smaller mean-square charge radius than their respective ground states irrespective of nuclear deformation. This offered an explanation into the observed odd-even staggering of nuclear charge radii~\cite{Zawischa1985} as arising from a combination of increasing rigidity (i.e. a reduction in the root mean square quadrupole deformation value towards the mean value) or decreasing surface diffuseness, due to the orbital (Pauli) blocking of the odd nucleon~\cite{Bissell2007}.

The multi-quasiparticle isomer studies looked into the effects of general nucleon pairing, not specifically $\pi\nu$ pairs, the existence of which has yet to be proved conclusively~\cite{Qi2015}, although many studies have yielded results supporting the idea. The premise for using odd-odd $N=Z$ nuclei as a testbed for $\pi\nu$ pairing correlations is supported by the investigation in $^{38}$K ($N,Z=19$) in which the relative difference in ground and isomeric state charge radii was probed via a direct measurement of the $^{38\mathrm{g},38\mathrm{m}}$K isomer shift~\cite{Bissell2014}. In this system, the $I=3$, $T=0$ ground state was found to have a smaller charge radius than the $I=0$, $T=1$ isomer, thus validating the prediction of this intuitive $\pi\nu$ pairing model. Shell model calculations appear to reproduce the effect quantitatively for $^{38\mathrm{g},38\mathrm{m}}$K~\cite{Bissell2014}. Such a remarkable agreement motivates the study of other $N=Z$ nuclei. Above the $N=Z=20$ shell closures, an inversion of the ground and isomeric states is observed and the $I=0^+$, $T=1$ becomes the ground state. The isomer shift has only been measured in one such system, $^{50\mathrm{g},50\mathrm{m}}$Mn~\cite{Charlwood2010a}, which shows an effect of similar magnitude as in $^{38}$K. In $^{42}$Sc a $I=0^+$, $T=1$ is the ground state, and a long-lived isomer is present with $I=7^+$, $T=0$~\cite{Chiara2007, Scholl2007}. The measurement of this $^{42\mathrm{g},42\mathrm{m}}$Sc isomer shift, where the ground state is expected to be larger, will test if this simple trend continues and whether it can be  quantitatively understood.

\section{Experimental methodology}

Collinear laser spectroscopy~\cite{Cheal2010,Campbell2016} was performed at the IGISOL IV facility~\cite{vormawah18a}, located at the University of Jyv\"{a}skyl\"{a} JYFL Accelerator Laboratory, Finland, where beams of short-lived radioactive ions can be produced~\cite{Moore2014}. Singly-charged $^{42}$Sc ions were produced via a 30\,MeV $^{40}$Ca($\alpha$,pn)$^{42}$Sc fusion-evaporation reaction, taking place in an environment of a helium buffer gas at a pressure of $\sim$150\,mbar inside the IGISOL chamber. The use of thin foil targets (thickness $\sim$2\,mg/cm$^{2}$) at IGISOL, coupled with the extraction of reaction products via a supersonic gas jet, enables fast release irrespective of physical or chemical properties. Once extracted from the target chamber, the reaction products were then formed into a 30\,keV beam of singly-charged ions which were subsequently mass separated using a 55$^{\circ}$ dipole magnet. Ions were cooled and bunched using a gas-filled radio-frequency quadrupole \cite{NieminenHF}, from which bunches of temporal width 20\,$\mu$s were released every 100\,ms and directed to the laser spectroscopy station. 

Ions were overlapped in an anti-collinear geometry with 0.5\,mW of light from a frequency-doubled continuous-wave narrow-linewidth dye laser, running with Pyridine 2 dye. Optical spectra were taken using the same 363.1\,nm $3d4s$ $^{3}D_{2} \rightarrow 3d4p$ $^{3}F_{3}$ atomic transition as used in a previous study of radioactive scandium isotopes~\cite{Avgoulea2010}. The fundamental frequency of the laser was locked to an I$_{2}$ absorption line and ions were Doppler tuned across the resonances by applying a voltage ramp to the photon-ion interaction region. For each voltage step an ion bunch was released and a gate was applied to the photon detection signal corresponding to the ion bunch transit time in front of the photomultiplier tube.
The resulting voltage scans were transformed into frequency, $f$, via
\begin{equation}
f=f_{L}(1+\alpha+\sqrt{2\alpha+\alpha^{2}}),
\end{equation}
where $\alpha=eV/mc^{2}$, $V$ is the total accelerating voltage and $f_{\rm L}$ is the laser frequency~\cite{Cheal2010}. Several independent scans of $^{42\mathrm{g},42\mathrm{m}}$Sc were taken and summed. The frequency conversion was performed using the atomic mass, $m$, of $^{42\mathrm{g}}$Sc~\cite{Wang2012}, and incorporating the 617\,keV excitation of the isomer~\cite{Chiara2007,Scholl2007} as appropriate.

\section{Data analysis and results}

\begin{figure}
 \includegraphics[width=\linewidth]{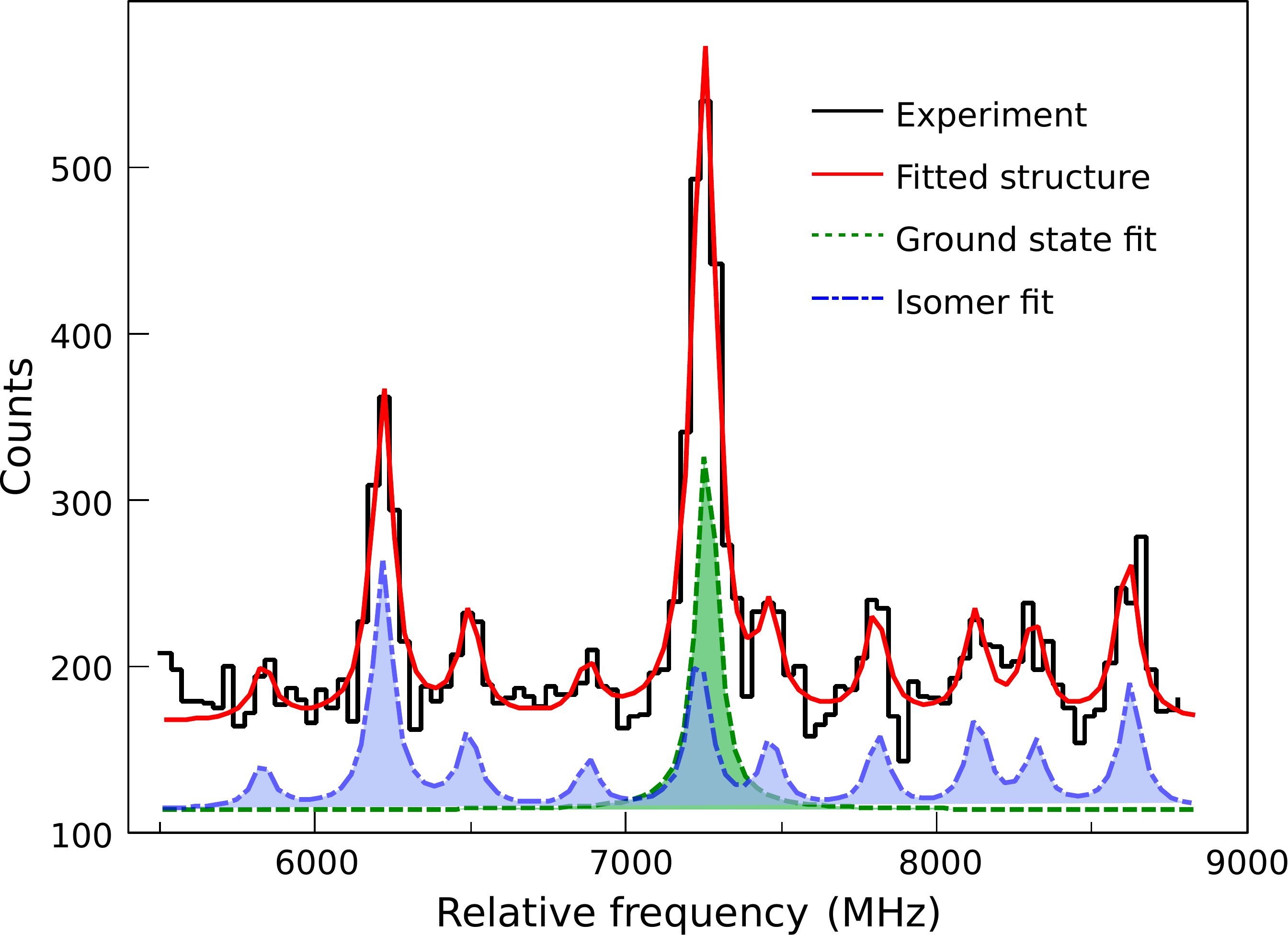}
 \caption{\label{fig:42ScSumplusfit}Measured and fitted hyperfine spectrum of $^{42}$Sc measured on the 363.1\,nm line, together with the separate contributions from the ground and isomeric states (offset for clarity).}
\end{figure}

A chi-squared minimisation routine was used in order to fit a series of Lorentzian peaks to the data, from which the hyperfine $A$ (magnetic dipole) and $B$ (electric quadrupole) parameters~\cite{Cheal2010} were extracted for the upper electronic state ($A_\mathrm{u}$, $B_\mathrm{u}$). Ratios of $A$ and $B$ parameters for the upper and lower states were constrained to ${A_{\mathrm{l}}}/{A_{\mathrm{u}}}=2.469$ and ${B_{\mathrm{l}}}/{B_{\mathrm{u}}}=0.552$~\cite{Avgoulea2010}. Frequencies of the isomeric peaks were therefore calculated according to
\begin{equation}
\gamma=\nu+(\alpha_{\rm u}-2.469\alpha_{\rm l})A_{\rm u}+(\beta_{\rm u}-0.552\beta_{\rm l})B_{\rm u}
\end{equation}
where, for each state,
\begin{eqnarray}
\alpha&=&\frac{K}{2},\\
\beta&=&\frac{3K(K+1)-4I(I+1)J(J+1)}{8I(2I-1)J(2J-1)},
\end{eqnarray}
with $K=F(F+1)-I(I+1)-J(J+1)$ and $\nu$ is the centroid of the isomer. Each peak was assigned a free intensity parameter, but since one isomer peak is obscured by the single ground-state peak, its intensity was constrained with respect to the most intense isomer peak assuming a model intensity distribution. Figure~\ref{fig:42ScSumplusfit} shows the summation of the $^{42}$Sc measurements. Hyperfine $A$ and $B$ coefficients and the isomer shift obtained from the fitting are shown in Table~\ref{tab:HyperfineParameters}.

\begin{table}
\caption{\label{tab:HyperfineParameters}Values of the hyperfine magnetic dipole, $A_{\rm u}(^{3}F_{3})$, and electric quadrupole, $B_{\rm u}(^{3}F_{3})$, parameters for $^{42\rm m}$Sc, measured on the 363.1\,nm line.\newline}
\small
\begin{tabular}{llll}
\hline
\hline
A&$A_{\mathrm{u}}$ (MHz) & $B_{\mathrm{u}}$ (MHz) & $\delta\nu^{42\rm g, m}$ (MHz) \\
\hline
 $^{42\rm m}$Sc&$+82.6(3)$ & $-34(28)$ & +74(5) \\
\hline
\hline
\end{tabular}
\end{table}

Using the values in Table~\ref{tab:HyperfineParameters} and the known hyperfine coefficients and nuclear moments of $^{45\mathrm{g}}$Sc~\cite{Avgoulea2010}, the magnetic dipole moment, $\mu$, and spectroscopic electric quadrupole moment, $Q_{\mathrm{s}}$, were calculated for $^{42\mathrm{m}}$Sc and are shown in Table~\ref{tab:42ScNuclearMoments}. 
\begin{table}
\caption{\label{tab:42ScNuclearMoments}Nuclear moments for $^{42\mathrm{m}}$Sc, calibrated using the published values for $^{45\mathrm{g}}$Sc~\cite{Avgoulea2010}. Also shown is the change in mean-square charge radius between the ground and isomeric state, $\delta\langle r^2\rangle^{42\rm g,42 \rm m}$, determined from the isomer shift using the revised calculation of $F=-349(15)~$MHz/fm$^{2}$, with the corresponding systematic error shown in square brackets.\newline}
\small
\begin{tabular}{llll}
\hline
\hline
A & $\mu$ ($\mu_{N}$) & $Q_{\mathrm{s}}$ (b) & $\delta\langle r^2\rangle^{42\rm g, m}$ (fm$^2$) \\\hline
$^{42m}$Sc & $+3.820(14)$ & $-0.12(10)$ & $-0.212(14)[9]$ \\
\hline
\hline
\end{tabular}
\end{table}

The change in mean-square charge radius between $^{42\mathrm{g}}$Sc and $^{42\mathrm{m}}$Sc was extracted from the $^{42\mathrm{g},42\mathrm{m}}$Sc isomer shift. The change in nuclear mean-square charge radius, $\delta\langle r^2\rangle^{A,A'}=\langle r^2\rangle^{A'}-\langle r^2\rangle^{A}$, is related to an isotope or isomer shift, $\delta\nu^{A,A'}=\nu^{A'}-\nu^A$, by~\cite{Cheal2012}
\begin{equation}{\label{IsotopeShift}}
\delta\nu^{A,A'}=M\frac{m_{A'}-m_A}{m_Am_{A'}} + F\delta \langle r^{2}\rangle^{A,A'},
\end{equation}
where $F$ and $M$ are the respective atomic factors for the field shift and mass shift. These are calculated using the multi-configuration Dirac-Fock (MCDF) method for a specific atomic transition, but are independent of the isotopes under study~\cite{Cheal2012}.

\begin{figure}
\includegraphics[width=0.9\linewidth]{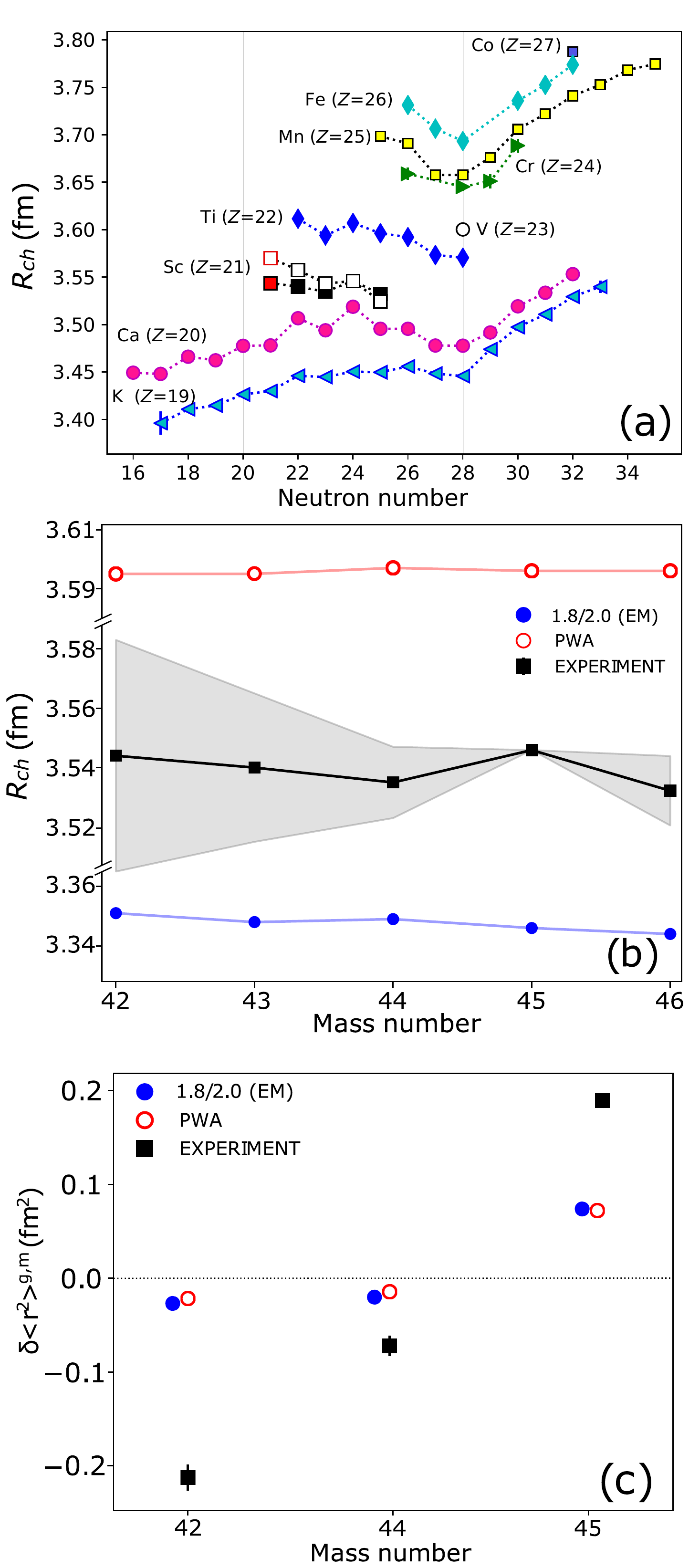}
\caption{\label{fig:careg} (a) Experimental nuclear charge radii in the calcium region~\cite{Angeli2013, kosz2020charge}. The full black squares show the charge radii of the Sc isotopes obtained using the newly calculated $F$ and $M$. For comparison, the literature values from~\cite{Avgoulea2010} are presented by empty squares. The new results for $^{42}$Sc are shown in red squares. The systematic uncertainties due to the atomic parameters are not shown. (b) The comparison of the measured charge radii to IMSRG calculations using two different interactions. The shaded area indicates the systematic uncertainty arising from the atomic parameters. (c) Measured and calculated changes of the mean-square charge radii of $^{42g,42m}$Sc, $^{44g,44m}$Sc and $^{45g,45m}$Sc. }
\end{figure}

Previous MCDF calculations yielded values of $F=-355(50)$\,MHz/fm$^{2}$ and $M=+583(30)$\,GHz$\cdot$u for the 363.1\,nm ${^{3}D_{2}} \rightarrow {^{3}F_{3}}$ transition in the Sc$^{+}$ ion~\cite{Avgoulea2010}. Revised calculations were performed as part of this work, in which the full relativistic recoil Hamiltonian was applied~\cite{Naze2013,FRITZSCHE20191}.  Two sets of calculations were performed. The first utilised a model similar to that employed in~\cite{Avgoulea2010}, and a multi-reference set with a configuration of $3d4s$, $3d^{2}$, $4s^{2}$, $4p^{2}$ for the ground state and $3d4p$ for the excited state. These calculations provided an estimate of the effect of the full relativistic recoil Hamiltonian but failed to reproduce the experimental value of the transition energy, hinting at the use of an unbalanced multi-reference set. A second set of calculations was hence performed using an expanded configuration of $3d4p$, $4s4p$ for the excited state and including double excitations from the $3s$ shell to account for additional core effects. The result of this was a much reduced uncertainty on $F$, whilst the total mass shift is taken from the MCDF calculations, rather than using the scaling law for the normal mass shift. 

For the 363.1\,nm transition used here and in~\cite{Avgoulea2010}, the newly calculated values of $F=-349(15)~$MHz/fm$^{2}$ and $M=+625(60)$~GHz$\cdot$u are adopted. A recalculation of the previously measured mean-square charge radii~\cite{Avgoulea2010} is shown in Table~\ref{tab:gs-radii}. While the field shift factor is similar to the previous value, an increase in the calculated mass shift factor produces a trend in the ground state radii more in keeping with regional systematics, as shown in Figure~\ref{fig:careg}(a). For changes in the mean-square charge radii in an isotopic chain $\delta \langle r^2 \rangle^{A,A'}$, systematic errors are dominated by the mass shift factor. This error is identically zero for the reference isotope, $^{45}$Sc. On the other hand, the changes in the mean-square charge radii of different states in the same isotope $\delta \langle r^2 \rangle^{g,m}$ are calculated with respect to the ground state, resulting in reduced systematic uncertainties. In addition, the mass shift (and its error) is negligible due to very small mass differences between ground and isomeric states. Therefore the value for $\delta\langle r^{2}\rangle^{42\mathrm{g},42\mathrm{m}}$, shown in Table~\ref{tab:gs-radii}, was calculated solely from the $F$ factor and is not affected by the error on $M$. When the nuclear charge radii $R_{ch}$ are calculated from the $\delta\langle r^{2}\rangle^{45,A'}$, the charge radii of the stable $^{45}$Sc is used as a reference \cite{Angeli2013}, thus the systematic uncertainties are propagated to all the other calculated values of the ground and isomeric states.  
\begin{table}
\caption{\label{tab:gs-radii}Changes in mean-square charge radius of the Sc isotopes, recalibrated using the newly calculated values of $F=-349(15)~$MHz/fm$^{2}$ and $M=+625(60)$~GHz$\cdot$u calculated as part of this work. \newline}
\small
\begin{tabular}{llllll}
\hline
\hline

 $A'$ &A   & $\delta\langle r^{2} \rangle^{A,A'}$ (fm$^{2}$) & $R_{ch}$ (fm)        
 \\\hline
 
 42 & 45   & $-0.016(31)[273]$   & $3.544(5)[39]$       \\
 42m&  42  & $-0.212(14)[9]$     & --  \\
 43 &  45 & $-0.042(14)[178]$  & 3.540(3)[25]           \\
 44 &  45  & $-0.081(11)[87]$    &$3.535(3)[12]$           \\
44m &  44  & $-0.072(11)[3]$    & -- \\
 45 &  45  & 0                  &3.5459(25)                  \\
45m &  45  & $+0.189(6)[8]$     & -- \\
 46 &  45  & $-0.097(9)[83]$     &$3.5322(28)[117]$            \\

\hline
\hline
\end{tabular}
\end{table}


\section{Discussion}
The measured electromagnetic moments of $^{42m}$Sc presented in Table \ref{tab:42ScNuclearMoments} can be used to better understand the nuclear structure of this isomer. The magnetic dipole moment gives an insight into the leading proton and neutron configuration of this state while the electric quadruple moment provides insight on the deformation. The $^{42}$Sc self-conjugate isotope is expected to have a rather simple nuclear structure given that it is made up of one proton and one neutron outside the magic $^{40}$Ca core. The nuclear spin of $I=7$ for $^{42\mathrm{m}}$Sc~\cite{Chiara2007} results in an empirical estimate~\cite{Neugart2006} of $\mu=+3.80$\ $\mu_{\rm N}$ for the magnetic dipole moment, from an average of the neighbouring $^{41}$Sc and $^{43}$Sc isotopes, coupled to an average of the $^{41}$Ca and $^{43}$Ti isotones~\cite{Stone2005}. This supports the stretched $[\pi f_{7/2} \otimes \nu f_{7/2}]\,7^+$ configuration. The effective quadrupole moment of $^{42m}$Sc was also calculated using the measured quadrupole moments of $^{41}$Ca and $^{41}$Sc yielding in $Q=$-0.211(18) b, somewhat larger than the measured $Q=$-0.12(10) b, but consistent given the large uncertainty of the latter. The measured electromagnetic moments of $^{42}$Sc are thus reproduced by simple empirical calculations, confirming that the properties of this state are dominated by the coupling of a $\pi$ and $\nu$ in the $f_{7/2}$ orbital. The small value of $Q$ and good agreement with the empirical estimate show that nuclear deformation is not expected to affect the size of this isomer.

A simple shell model approach had success in calculating the isomer shift between the $T=0$ and $T=1$ states of $^{38}$K. In the shell model, changes in mean-square charge radius can be calculated from the difference in proton occupancies of the $fp$ shell in the ground state and the isomer using equation~\cite{Bissell2014}

\begin{equation}{\label{ShellModelChargeRadii}}
\small
\delta\langle r^{2}\rangle^{A,A'}=\frac{1}{Z}\Delta n^{\pi}_{fp}(A,A')b^{2},
\end{equation}
where $b$ is the oscillator parameter, and $\Delta n^{\pi}_{fp}$ is the change in proton occupancy of the $fp$ shell between two isotopes or nuclear states $A$ and $A'$. To determine $b^{2}$, the equation of Duflo and Zuker~\cite{Duflo1999} was used:
\begin{equation}{\label{OscillatorParameterSquared}}
\small
b^{2}=1.07A^{\frac{1}{3}}\bigg\lgroup 1-\bigg\lgroup\frac{2T}{A}\bigg\rgroup^{2}\bigg\rgroup e^{\frac{3.5}{A}},
\end{equation}
where $T$ stands for the isospin, yielding a value of $b^{2}=4.043$\,fm$^{2}$ for $^{42}$Sc. 
\begin{table}
\caption{\label{tab:ScShellModelNuclearMoments}Empirical estimates of the nuclear magnetic dipole and electric quadrupole moments for $^{42\mathrm{m}}$Sc together with those calculated in the shell model~\cite{Caurier2001} with the ZBM2 and ZBM2M interactions \cite{Bissell2014} and the measured value.\newline}
\small
\begin{tabular}{lllll} 
\hline
\hline
 & Empirical & ZBM2 & ZBM2M & Experiment  \\\hline
$\mu$ $(\mu_{N})$ & +3.8 &$+3.878$ & $+3.878$& +3.820(14) \\
$Q_{\mathrm{s}}$ (b) & -0.211(18) &-0.165 & $-0.178$ & -0.12(10) \\
 \\
\hline
\hline
\end{tabular}
\end{table}
To determine the proton occupancy, shell model calculations were performed in the model space consisting of the $s_{1/2}$, $d_{3/2}$, $f_{7/2}$ and $p_{3/2}$ orbitals for both protons and neutrons above an inert $^{28}$Si core, using the shell model code NuShellX \cite{Brow14Nush}, for which full diagonalisation of this model space has been achieved. Two sets of calculations were performed; one using the original ZBM2 interaction, and one using the version with modified $V^{0,1}_{d_{3/2}d_{3/2}}$ matrix elements. The same methodology was used for the calculations of $^{38}$K~\cite{Bissell2014}, where the latter interaction was found to reproduce the properties of this self-conjugate isotope more accurately. Magnetic dipole moments (using free $g$-factors) and electric quadrupole moments calculated from the wave functions are presented in Table~\ref{tab:ScShellModelNuclearMoments} and show good agreement with the experimental values.

Table~\ref{tab:42mradii} shows a comparison between the experimental value of $\delta\langle r^{2}\rangle^{42\mathrm{g},42\mathrm{m}}$ and the theoretical calculation using the shell model via equation~\ref{ShellModelChargeRadii}. Unlike the case of $^{38}$K where a close match is seen between the shell model calculation and the experimental value, for $^{42}$Sc the change in mean-square charge radius is underestimated by a factor of two.

\begin{table}
\caption{\label{tab:42mradii} Changes in the mean-square charge radii of the ground state and isomer in $^{\mathrm{42}}$Sc expressed in fm$^2$. The experimentally measured value is compared to calculations from the shell mode using the ZBM2 and ZBM2M interactions and to the results of the IMSRG method using the PWA and 1.8/2.0(EM) interactions.  \newline}
\small
\begin{tabular}{llllll} 
\hline
\hline
ZBM2  & ZBM2M & PWA & 1.8/2.0(EM) & Experiment \\\hline
-0.114 & -0.099 & -0.022 & -0.027 & -0.212(14)[9] \\\hline

\hline
\hline
\end{tabular}
\end{table}

Ab initio calculations were performed using VS-IMSRG predictions for the charge radii of light scandium isotopes using two initial sets of NN+3N forces from chiral effective field theory~\cite{Epel09RMP,Mach11PR}. The VS-IMSRG, developed over Refs.~\cite{Tsuk12SM,Bogn14SM,Morr15Magnus,Stro16TNO,Herg16PR,Parz17Trans,Stro17ENO,Stro19ARNPS} decouples a valence-space Hamiltonian and consistent operators from the full Hilbert space via an approximate unitary transformation. Here, to provide some assessment of uncertainty from starting nuclear forces, we use both the 1.8/2.0(EM) and PWA from a well-established family of chiral interactions~\cite{Hebe11fits,Simo16unc,Simo17SatFinNuc}. The former reproduces ground-state and excitation energies throughout the medium- to heavy-mass region, including nuclear driplines~\cite{Holt19drip}, but generally underpredicts absolute charge radii, while PWA gives larger radii but underbinds finite nuclei~\cite{Ruiz16Ca52,Simo17SatFinNuc,Morr17Tin}. Recently, both were shown to reproduce the odd-even staggering in copper isotopes well~\cite{Groo20Cu}. To obtain charge radii, we first decouple the core and valence-space intrinsic proton mean-squared radius operator, then apply corrections arising from the mean-square charge radii of the proton and the neutron, as well as the relativistic Darwin-Foldy and spin-orbit corrections, as detailed in Ref.~\cite{Ruiz16Ca52}.

We use the IMSRG++ code~\cite{Stro17imsrg++}, in the IMSRG(2) approximation where induced many-body operators are truncated at the two-body level. In addition, the capability to generate valence-space Hamiltonians across major harmonic-oscillator shells was developed in Ref.~\cite{Miyagi2020}. In the current work, we take $^{28}$Si as the core and decouple a valence-space Hamiltonian for the space spanned by both proton and neutron $1s_{1/2}$, $0d_{3/2}$, $0f_{7/2}$, and $1p_{3/2}$ orbitals unique for each isotope studied. The resulting valence-space Hamiltonians are diagonalized with the NuShellX@MSU shell-model code~\cite{Brow14Nush} (and the KSHELL code in some cases~\cite{Shimizu2019}) to obtain ground-state energies and expectation values for the intrinsic proton mean-square radius operator. 

We start from a harmonic-oscillator basis of 15 major shells (i.e., $e=2n+l \leqslant e_{\mathrm{max}}=14$) at $\hbar\omega=16$~MeV then transform to the Hartree-Fock basis, capturing the effects of 3N forces among valence nucleons with the ensemble normal ordering described in Ref.~\cite{Stro17ENO}. 
Using the approximate unitary transformation from the Magnus framework we additionally decouple a valence-space radius operator consistent with the valence-space Hamiltonian.
In addition, for storage requirements, we impose a cut of $e_{1}+e_{2}+e_{3} \leqslant E_{\mathrm{3Max}}=16$ for 3N matrix elements. 
Finally, spurious center-of-mass modes are separated by adding the center-of-mass Hamiltonian with the coefficient $\beta$ at the beginning of the calculation as descussed in Ref.~\cite{Miyagi2020}. The $\beta$-dependence of the results is small around $\beta=3$, and thus the following discussion is based on the results with $\beta=3$.

\begin{figure}
\includegraphics[width=\linewidth]{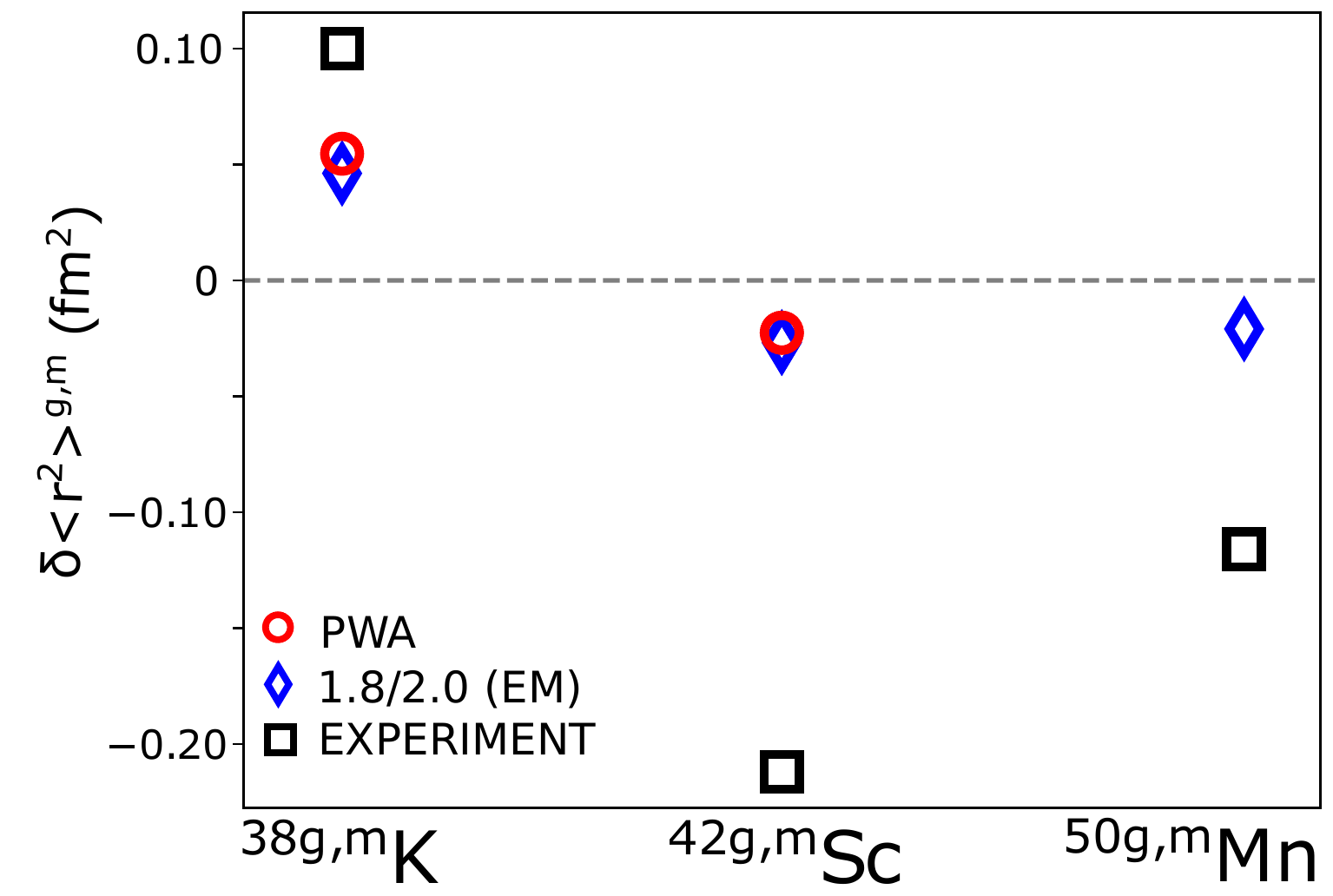}
\caption{\label{fig:all_NZ_istopes}Experimental isomer shifts of the $T=0$ and $T=1$ states in $N=Z$ isotopes of K, Sc and Mn together with the theoretical values calculated with the IMSRG method. The error bars are smaller than the markers. }
\end{figure}

The comparison of the measured and calculated charge radii using the VS-MSRG is shown in Fig. \ref{fig:careg}(b). As expected, results using the PWA interaction overestimate the charge radii, while the 1.8/2.0(EM) gives too-small values. While the flat gradient of the trend seen in the scandium chain is overall well reproduced, the details are not consistent with the experimental data. 
Next the changes in the mean-square charge radii of the ground state and isomer of $^{42g,42m}$Sc, $^{44g,44m}$Sc and $^{45g,45m}$Sc are shown in \ref{fig:careg}(c). 
Note that the systematic error of the experimental values are reduced because the corresponding ground state is chosen as the reference state. 
The systematic errors are smaller than the statistical uncertainties for the ground-state isomer-pairs in $^{42,44}$Sc and are of similar magnitude for $^{45}$Sc, as shown in the third column of Table \ref{tab:gs-radii}. 
The sign of the isomer shifts is correctly reproduced in all the measured cases in the Sc chain. However, the magnitude is underestimated. For $^{42m}$Sc in particular, the change in mean-square charge radius with respect to the ground state is underestimated by a factor of 10, as shown in Table~\ref{tab:42mradii}.
This significant underestimation can be partially understood by looking into the occupancy in the ground and isomer states since the radius operator is dominated by the one-body piece.
We observed that number of protons excited to the $pf$ shell are $\sim 0.6$ and $\sim 0.5$ in the ground and isomer states, respectively, and thus $\Delta n^{\pi}_{fp} \sim 0.1$, considerably smaller than the $\sim 0.5$ from the ZBM2 calculations.
Less excitation to $pf$ orbitals indicates the overestimation of the single-particle gap between $sd$ and $pf$ shells, which would be due to the IMSRG(2) approximation as seen in comparison with the coupled-cluster method~\cite{Taniuchi2019}.
Also, we observed that these fewer proton excitations from $sd$ to $pf$ could not solely explain the underestimation of the isomer shift. 
For example, even if we assume moderate proton one-particle-one-hole excitation from $d_{3/2}$ to $f_{7/2}$ in the ground state, the corresponding isomer shift is $-0.165$ fm$^{2}$, which is still insufficient to explain the experimental isomer shift.
We would need to include additional physics such as the radius operator renormalization effect, or the implementation of IMSRG(3)-level approximation, which is currently underway.

\begin{figure}
\includegraphics[width=\linewidth]{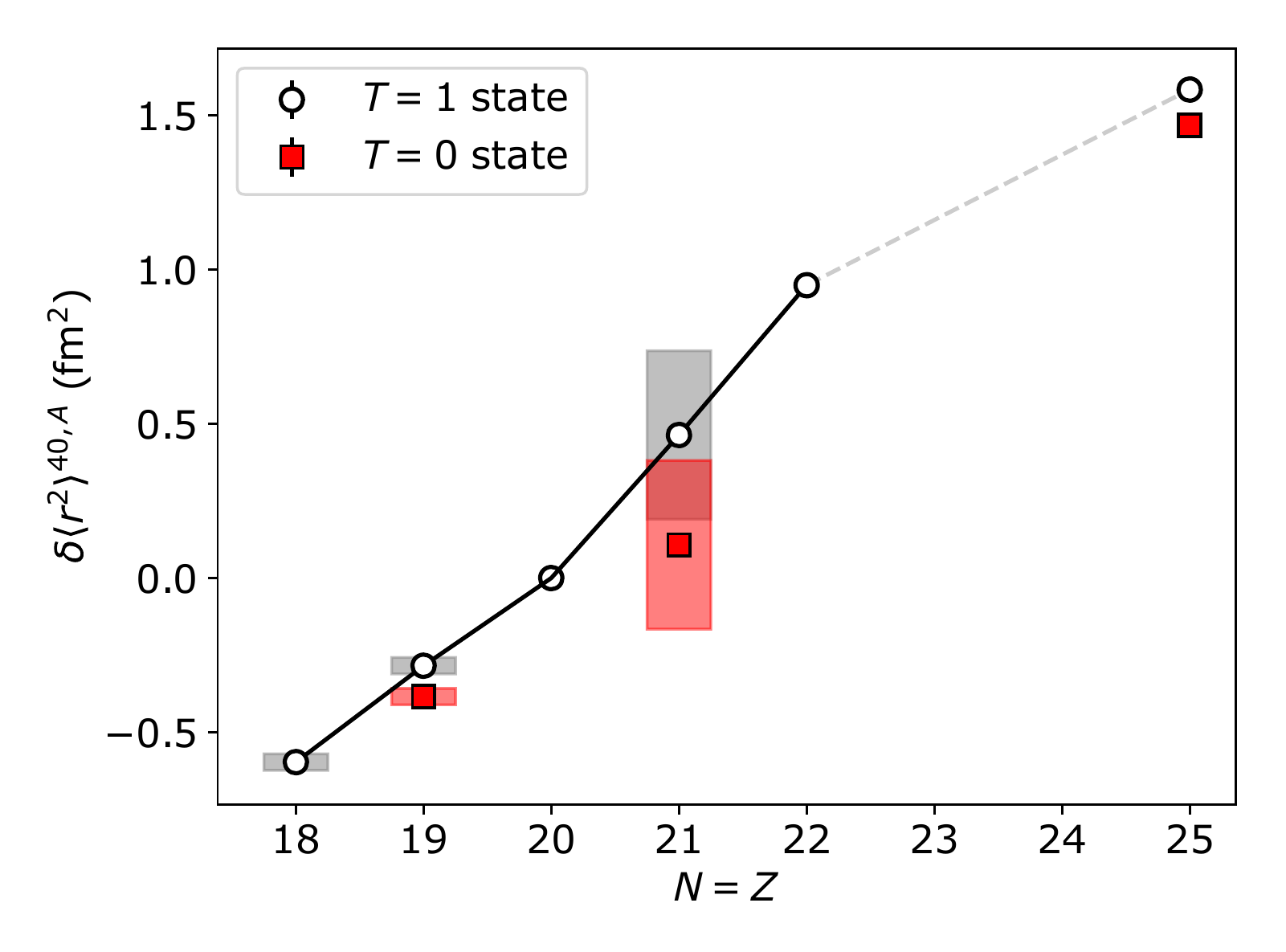}
\caption{\label{fig:NeqZplot}Changes in mean-square charge radius of self-conjugate nuclei from $^{36}$Ar to $^{50}$Mn, relative to $^{40}$Ca. In each case, absolute values of $\langle r^2\rangle$ for reference isotopes~\cite{Angeli2013} are added to isotope shifts accordingly~\cite{Bissell2014,Avgoulea2010,Angeli2013}. The shaded area corresponds to the systematic error due to the uncertainty of the atomic parameters.\newline}
\end{figure}

Taking advantage of the universal applicability of the VS-IMSRG method, the difference in size between the $T=0$ and $T=1$ states of the $Z=N$ isotopes for K, Sc and Mn were also calculated with the same interactions. These results are compared to  experimental values in Fig.~\ref{fig:all_NZ_istopes}. 
The sign of these differences is correctly predicted, despite changing from the $sd$ shell $^{38}$K isotope to the $pf$ shell $^{42}$Sc and $^{50}$Mn. 
The magnitude, however, is clearly underestimated. 
Finally, Fig.~\ref{fig:NeqZplot} shows the variation of absolute mean-square charge radii along $N = Z$ from $^{36}$Ar to $^{50}$Mn. 
It is interesting to remark that the difference in size between the $T = 0$ and $T = 1$ states is almost equal in $^{38}$K and $^{50}$Mn, while it is twice as large in $^{42}$Sc. 
In the $\pi\nu$ pair blocking picture introduced earlier~\cite{Bissell2014}, the $T = 1$ state in an odd-odd nucleus should lie exactly on the $N = Z$ line established by the $T = 1$ ground states in the even-even neighbours, while the $T = 0$ state is smaller due to the blocking of pair scattering, as illustrated by the $^{38}$K case. 
From the sign of the $^{42\mathrm{g},42\mathrm{m}}$Sc isomer shift, it is immediately clear that the charge radius of the $0^+$  ground state is indeed greater than the charge radius of the $7^+$ isomer, verifying qualitatively this prediction. 
However, an unambiguous comparison with the neighbouring $T = 1$ states is hindered due to a large systematic uncertainty on the determination of the \emph{absolute} charge radii of $^{42\mathrm{g}}$Sc and $^{42\mathrm{m}}$Sc which arises from the uncertainty on $\langle r^2\rangle$ relative to the reference isotope $^{45}$Sc~\cite{Angeli2013} and therefore affects both states equally. 
Despite this large systematic uncertainty, our result is fully consistent with the $\pi\nu$ pairing picture.

\section{Conclusions}

The results which we report here, together with those of the previous study~\cite{Bissell2014}, are qualitatively consistent with the intuitive picture of $\pi\nu$ pairing correlations along the line of $N=Z$. The negative value obtained for $\delta\langle r^{2}\rangle^{42\mathrm{g},42\mathrm{m}}$ indicates a larger charge radius for the $0^{+}$ ground state than for the $7^{+}$ isomer, as expected from employing an orbital blocking picture. 

Quantitatively, on the other hand, the size of the phenomenon is not yet understood and requires further investigation both from the theoretical as from the experimental point of view. While the electromagnetic moments are well reproduced by the shell model  and simple empirical calculations, the mechanism determining the difference in the charge radius of the $T=1$ and $T=0$ states of the self-conjugate $^{42}$Sc is still missing. This is evidenced by both the shell model and \mbox{IMSRG} calculations. On the experimental side, the difference in charge radius of these two states in odd-odd self-conjugate nuclei is only known in two other cases besides $^{42}$Sc. A more complete picture of the evolution of such pairing correlations could be achieved by measuring the isomer shifts in other odd-odd self-conjugate nuclei over the nuclear chart. One ideal candidate for a future study would be $^{26}$Al, which has a $5^{+}$ ground state and a $0^{+}$ isomer. It is therefore the isomer in $^{26}$Al for which the charge radius is expected to be greater. Another such candidate would be $^{46}$V, in which the valence $\pi\nu$ pair once again occurs in the $f_{7/2}$ shell, and further cases such as $^{52}$Fe, $^{54}$Co and $^{94}$Ag, providing fertile ground for this new avenue of research.

\section{Acknowledgments}

We are grateful to J. Simonis for providing the 1.8/2.0 (EM) and PWA 3N matrix element files and S. R. Stroberg for the imsrg++ code~\cite{Stro17imsrg++} used to perform these calculations.
TRIUMF receives funding via a contribution through the National Research Council of Canada. This work was further supported by NSERC, the Arthur B. McDonald Canadian Astroparticle Physics Research Institute, Canadian Institute for Nuclear Physics. This work has been supported by the Academy of Finland under the Finnish Centre of Excellence Programme 2012-2017 (Project No. 251353), Nuclear and Accelerator Based Physics Research at JYFL), the UK Science and Technology Facilities Council (STFC). This work was also supported by the Bundesministerium
für Bildung und Forschung (BMBF, Germany) under Projects No. 05P18SJCIA.

\biboptions{sort&compress}
\bibliographystyle{model1-num-names}
\bibliography{refs.bib}

\end{document}